\documentstyle[prl,aps,multicol,epsfig,amsmath,amssymb]{revtex}

\begin{document}
\normalsize
\draft
\widetext

\title{
Time-dependent energy absorption changes during ultrafast lattice
deformation}
\author{H. O. Jeschke, M. E. Garcia and K. H. Bennemann}
\address{Institut f{\"u}r Theoretische Physik der Freien
Universit{\"a}t Berlin,
Arnimallee 14, 14195 Berlin, Germany,}
\date{\today}
\maketitle

\begin{abstract}
The ultrafast time-dependence of the energy absorption of covalent
solids upon excitation with femtosecond laser pulses is theoretically
analyzed. We use a microscopic theory to describe laser induced
structural changes and their influence on the electronic
properties. We show that from the time evolution of the energy
absorbed by the system important information on the electronic and
atomic structure during ultrafast phase transitions can be gained. Our
results reflect how structural changes affect the capability of the
system to absorb external energy.
\end{abstract}

\begin{multicols}{2}

\section{Introduction.}

For systems in thermodynamical equilibrium the occurrence of phase
transitions can be inferred from a change in the behavior of the
thermodynamic potentials (or its derivatives) as a function of an
intensive variable. For instance, divergences or abrupt changes in the
heat capacity $C=dU/dT$ are indicators for most first-order phase
transitions. In systems subject to extreme nonequilibrium conditions
like excitation with an intense femtosecond laser, neither
thermodynamic functions nor intensive variables are well
defined. However, a large variety of nonequilibrium phase transitions
can be induced by femtosecond laser pulses, as has been demonstrated
both experimentally and
theoretically~\cite{sokolowski:95,solis:98,jeschke:99}. Such ultrafast
phase transitions are detected indirectly by time-dependent changes of
the optical properties~\cite{downer:92,soko:98,mazur:01} or directly
by {\it post mortem} analysis~\cite{kautek:98} and recently through
more direct measurements using ultrashort X-ray
pulses~\cite{rose:99,rousse:01}. In a recent experimental study,
Musella {\it et$\:$al.} reported results on the excitation of graphite
with very long pulses (in the millisecond range). The thermogram
recorded during the heating of the sample shows a sharp increase of
the temperature slope at the melting point, which corresponds to
changes in the energy absorption $U_{\rm abs}(t)$ and heat
conductivity upon melting.  Note that in these experiments, thermal
equilibrium is reached during the excitation. Note that in the case of
femtosecond excitation $E_{\rm abs}(t)$ is a physical quantity,
whereas $C$ and $T$ are no longer well defined. The question which
immediately arises from this analysis is: Does $E_{\rm abs}(t)$ show
similar changes as $U_{\rm abs}$ in the case of thermal equilibrium
when the system is illuminated by a femtosecond laser pulse?

In this paper we show that this is the case, and that $E_{\rm abs}$
exhibits dramatic changes as a function of laser intensity $I$ when
the system undergoes an ultrafast phase transition. This leads to a
natural generalization of ``heat capacity'' for laser induced
nonequilibrium processes as $C=dE_{\rm abs}/dI$.

In Sec.~\ref{sec:theory} we briefly review our method of
calculating the structural response of covalent materials during and
after intense laser excitation. In Sec.~\ref{sec:results} we present
results for diamond and graphite, concentrating on the efficiency with
which energy is absorbed from the laser pulse. We show that the
structural changes during the action of the pulse are particularly
important for the understanding of the nonlinear dependence of the
absorbed energy on the intensity of the pulse. As these structural
changes occur on a time scale of approximately 100~fs the pulse
duration is responsible for very different absorption behavior in the
$\tau=20$~fs to $\tau=500$~fs pulse duration range. Finally, in
Sec.~\ref{sec:summary} we summarize our results.

\section{Theory.}\label{sec:theory}

The calculation of the energy absorption of covalent solids is based
upon a theory for the analysis of laser induced ultrafast processes in
solids. The theory employs a molecular dynamics method on the basis of
an electronic tight-binding Hamiltonian. This real-space calculation
takes into account all atomic degrees of freedom. Special attention is
paid to the strong nonequilibrium created in the electronic system by
the ultrashort laser pulse. A method of calculating nonequilibrium
occupation numbers for the energy levels of the system leads to a
molecular dynamics calculation on time-dependent potential energy
surfaces. This approach provides a theoretical framework for the
treatment of strong nonequilibrium situations in materials where
atomic and electronic degrees of freedom play an equally important
role.

In this theory trajectories for laser induced structural changes are
determined by molecular dynamics simulations for which we calculate
forces from the Hamiltonian
\begin{equation}\label{eq:hamop}
  H =  H_{\rm TB} + \sum_{i<j} \phi (r_{ij})\,.
\end{equation}
The second term contains a repulsive potential $\phi(r_{ij})$ that takes
care of the repulsion between the ionic cores, and the first term is a
tight-binding Hamiltonian
\begin{equation}\label{eq:hamtb}
H_{\rm TB} = \sum_{i\eta} \epsilon_{i\eta} n_{i\eta} + \sum_{\substack{ij\eta\vartheta \\ j\not=i}}
t_{ij}^{\eta \vartheta} c_{i\eta}^+ c_{j\vartheta}^{ }\,.
\end{equation}
Here, the first term is the on-site contribution. In the second term,
the $t_{ij}^{\eta \vartheta}$ are the hopping integrals, and $c_{i\eta}^+$ and
$c_{j\vartheta}^{ }$ are creation and annihilation operators for an electron
at site $i$ or $j$ in orbitals $\eta$ or $\vartheta$, respectively.

From the Hamiltonian, Eq.~(\ref{eq:hamop}), we calculate forces using
the Hellman-Feynman theorem:
\begin{equation}\begin{split}\label{eq:forces}
  {\bf f}_k(\{r_{ij}(t)\},t) = -& \sum_{m} n(\epsilon_m,t) \langle m|\nabla_k\: H_{\rm TB}(\{r_{ij}(t)\}) |m \rangle 
   \\-& \sum_{i<j} \nabla_k\:\phi (r_{ij}) \,.
\end{split}\end{equation}
Here, $|m \rangle$ stands for the eigenvector of the Hamiltonian $H_{\rm
TB}$ that corresponds to the eigenvalue $\epsilon_m$. The special feature of
the theory which makes it applicable to optically excited materials is
contained in the time dependent occupation numbers $n(\epsilon_m,t)$ for the
energy levels $\epsilon_m$ of the system. While in thermal equilibrium these
occupation numbers are calculated from a Fermi-Dirac distribution
function $n^0(\epsilon_m) = 2/(1+\exp{\{(\epsilon_m-\mu)/k_{\rm B}T_{\rm
e}\}})$ at a given electronic temperature
$T_{\rm e}$, electronic nonequilibrium is accounted for by solving
equations of motion for the occupation of electronic states:
\begin{equation}\begin{split}\label{eq:absorpthermal}
\frac{dn(\epsilon_m,t)}{dt} =& \int_{-\infty}^{\infty} d\omega\; 
g(\omega,t-\Delta t) 
\biggl\{\left[ n(\epsilon_m- \hbar \omega, t-\Delta t)  \right.  \\ 
& \left.  + n(\epsilon_m+ \hbar \omega, t-\Delta t)  - 2n(\epsilon_m, 
t-\Delta t)\right] \biggr\}  \\  
 & - \frac{n(\epsilon_m,t) - n^0(\epsilon_m,T_{\rm e})}{\tau_1} \,.
\end{split}\end{equation}
Thus, the electronic distribution is at each time step folded with the
pulse intensity function $g(\omega,t)$. This means that at each time step,
the occupation of an energy level $\epsilon_m$ changes in proportion to the
occupation difference with respect to levels at $\epsilon_m - \hbar \omega$ and at
$\epsilon_m + \hbar \omega$. In Eq.~(\ref{eq:absorpthermal}), constant optical matrix
elements are assumed. The second term of Eq.~(\ref{eq:absorpthermal})
describes the electron-electron collisions that lead to an
equilibration of the electronic system with a rate equation of the
Boltzmann type for the distribution $n(\epsilon_m,t)$. Hence, with a time
constant $\tau_1$, the distribution $n(\epsilon_m,t)$ approaches a Fermi-Dirac
distribution $n^0(\epsilon_m)$.

The electron temperature $T_{\rm e}$ that results from the electron
thermalization will not remain constant over time, but will decrease
due to electron-phonon coupling and to diffusion of hot electrons out
of the laser excited region of the solid into colder areas. In the
regime of an excitation of typically 10~\% of the valence electrons
into the conduction band for which this theory is intended no precise
knowledge about thermal conductivity of the electrons and
electron-phonon coupling constants is available. Thus, a relaxation
time $\tau_2$ which characterizes the time scale of the total decrease of
the electron temperature by both the electron-phonon and by the hot
electron diffusion processes is assumed:
\begin{equation}\label{eq:ephloss}
\frac{d T_{\rm e}(t)}{d t} = - \frac{T_{\rm e}(t) - T(t)}{\tau_2}\,.
\end{equation}
This simple relaxation time approach has the advantage that the time
scale $\tau_2$ on which the electron-lattice equilibration takes place
can be taken from experiment.

The forces given by Eq.~(\ref{eq:forces}) can now be used to solve the
equations of motion for the atoms numerically. In the case of a bulk
system, a constant pressure molecular dynamics (MD) scheme is
used. This is based on a Lagrangian which contains the shape and size
of the molecular dynamics supercell as additional degrees of
freedom~\cite{parah:80}:
\begin{equation}\label{eq:lagrange}
L = \sum_{i=1}^N \frac{m_i}{2} \dot{{\bf s}}^{\rm T}_i h^{\rm T} h\,
\dot{{\bf s}}_i + K_{\rm cell} - \Phi(\{r_{ij}\},t)- U_{\rm cell} \,.
\end{equation}
The first term is the kinetic energy of the atoms, with the
coordinates of the atoms ${\bf r}_i = h {\bf s}_i$ written in
terms of the relative coordinates ${\bf s}_i$ and the $3\times3$-matrix
$h$ that contains the vectors spanning the MD supercell;
$\dot{{\bf s}}_i$ are the relative velocity vectors, and T denotes
transposition. $\Phi(\{r_{ij}\},t)$ is the potential which is calculated
from a tight-binding formalism as explained above. The terms in the
Lagrangian of Eq.~(\ref{eq:lagrange}) which are responsible for the
simulation of constant pressure are an additional kinetic energy term
$K_{\rm cell}$ for which the simplest form is $K_{\rm cell} =
\frac{w_{\rm cell}}{2}\: {\rm Tr}(\dot{h}^{\rm T}\dot{h})$~\cite{parah:80}, and 
$U_{\rm cell}$ is an additional potential term which describes the
effect of an isotropic external pressure $U_{\rm cell} = p_{\rm
ext}\Omega$; $\Omega = \det(h)$ is the volume of the MD supercell.  The
equations of motion are derived from Eq.~(\ref{eq:lagrange}) by the
Euler-Lagrange formalism, and they are integrated numerically with the
velocity form of the Verlet algorithm~\cite{verlet:67,haile:92}.

\section{Results.}\label{sec:results}

The dependence of the absorption of energy on the laser pulse
intensity and duration shows an interesting behaviour. The following
results have been calculated in the bulk and for constant pressure.
Diamond and graphite samples of $N=64$ atoms per MD supercell were
thermalized to a temperature of $T = 300$~K by simulated annealing
before absorbing laser pulses of a wide range of intensities and
durations.

\begin{figure}
\hspace*{-0.2cm}\includegraphics[width=0.47\textwidth]{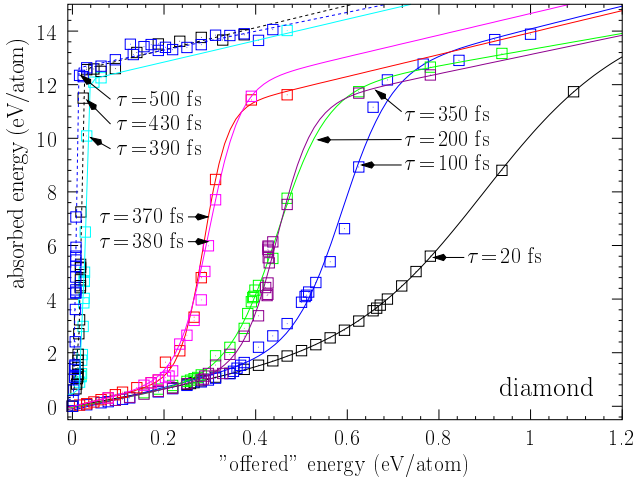}\\
\caption{
Absorbed energy per atom as a function of pulse intensity (``offered''
energy) for diamond. The pulse duration ranges from $\tau = 20$~fs to $\tau
= 500$~fs. The shape of the absorption curves and their ordering with
respect to the pulse duration can be understood from the
time-dependent changes of the DOS. Each of the results ($\boxdot$)
corresponds to a calculated trajectory. }
\label{fig:diamond_absorption}\end{figure}

Results for absorption of carbon in its diamond structure are shown in
Fig.~\ref{fig:diamond_absorption}. Every square ($\boxdot$)
corresponds to a calculated trajectory. Note, the lines connecting the
dots are drawn to guide the eye. The graph represents the actual
absorbed energy in units of electron volts per atom as a function of
the laser intensity for different pulse durations. This quantity is
denoted here as the ``offered'' energy. It is the energy that would have
been absorbed by the system, if no structural changes had taken
place. Thus, the ``offered'' energy is determined by performing the
calculation of energy absorption for fixed atomic coordinates. Note,
structural distortions change the electronic structure of the material
and consequently the absorption characteristics.

A general trend shared by all but the $\tau=390$~fs to $\tau=500$~fs curves
in Fig.~\ref{fig:diamond_absorption} is a moderate, linear increase of
absorption with intensity up to an absorbed energy of roughly
$E_0=1$~eV/atom. Then a sharp increase of absorption until
approximately $E_0=12$~eV/atom occurs. Then we find again a moderate
linear increase of absorption with laser intensity. The results show
that the absorption increases more sharply with increasing pulse
duration. With increasing duration of the laser pulse, the laser
intensity for which the sharp increase in absorption occurs decreases
significantly. The curves for long pulse durations are even
characterized by the absence of an initial weak increase of absorption
with intensity.

\begin{figure}
\hspace*{-0.2cm}\includegraphics[width=0.47\textwidth]{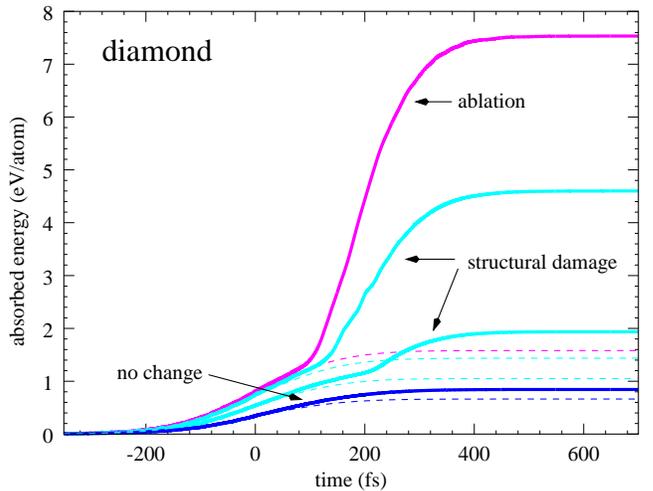}\\
\caption{
Absorbed energy per atom in diamond as a function of time for a pulse
duration of $\tau = 350$~fs. Absorption of trajectories with different
final states are shown. The dashed lines correspond to the absorption
that is expected if no structural changes occur during the action of
the pulse. }
\label{fig:d64TN350_absorption}\end{figure}

In Fig.~\ref{fig:d64TN350_absorption} we show the absorbed energy of
diamond as a function of time. The laser pulse duration was $\tau =
350$~fs. This figure serves as an example to clarify how the form of
the absorbed energy curves of Fig.~\ref{fig:diamond_absorption} comes
about. Fig.~\ref{fig:d64TN350_absorption} is directly related to the
$\tau = 350$~fs curve of Fig.~\ref{fig:diamond_absorption}: Each final
absorbed energy at $t = 700$~fs in Fig.~\ref{fig:d64TN350_absorption}
corresponds to a result ($\boxdot$) in the $\tau = 350$~fs curve of
Fig.~\ref{fig:diamond_absorption}.

Fig.~\ref{fig:d64TN350_absorption} shows the absorbed energy for four
different intensities. The final state of the material is
indicated. The lowest pulse intensity does not lead to structure
changes, due to the two medium intense pulses, the diamond structure
is damaged, and the strongest pulse leads to ablation of the
material. 
In Fig.~\ref{fig:d64TN350_absorption} the absorption that would have
been expected if no structural changes had occurred during the action
of the laser pulse is indicated by dashed lines. We observe that the
deviation of the energy actually absorbed by the material from the
expected value increases with increasing pulse intensity. In the case
of the trajectory where diamond remains unchanged, the actual
absorption is very close to the expected value. However, in
trajectories where structural damage or even ablation is observed the
actual absorption sharply deviates from the expected absorption.

\begin{figure}
\hspace*{-0.2cm}\includegraphics[width=0.47\textwidth]{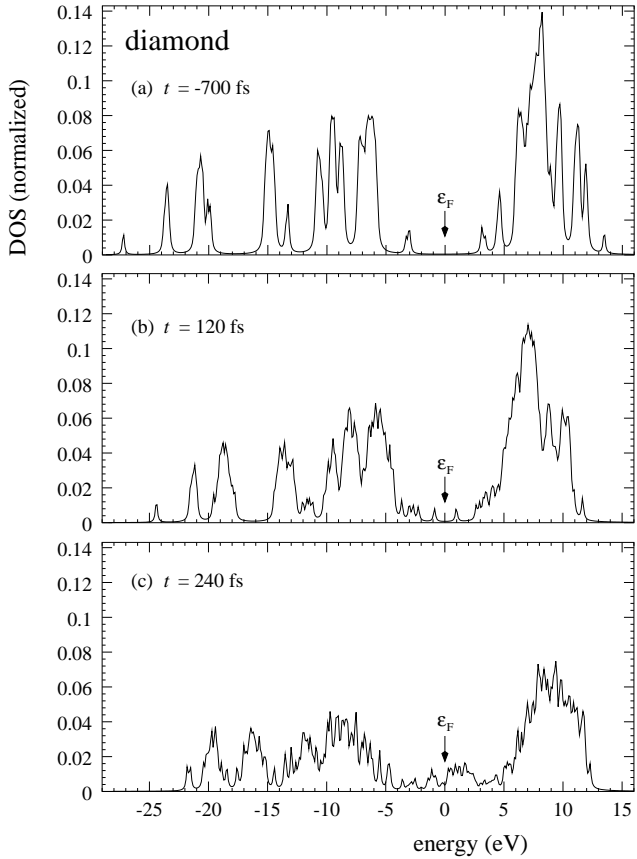}\\
\caption{
Density of states for diamond at different times for a pulse duration
of $\tau = 350$~fs. This figure corresponds to the trajectory with the
second highest absorbed energy of
Fig.~\ref{fig:d64TN350_absorption}. For a discussion of the features
of the DOS, see the text. }
\label{fig:d64TN350z.dos}\end{figure}

The cause for this sharp deviation will be explained with the help of
the densities of states shown in Fig.~\ref{fig:d64TN350z.dos}. They
correspond to the curve of Fig.~\ref{fig:d64TN350_absorption} with the
second-highest absorbed energy. In Fig.~\ref{fig:d64TN350z.dos}~(a)
the DOS before the laser pulse at $t = -700$~fs is shown; the
time is measured with respect to the maximum of the laser pulse of
$\tau=350$~fs duration. The DOS corresponds to diamond at $T =
300$~K. The DOS in Fig.~\ref{fig:d64TN350z.dos}~(b) is taken at $t
= 120$~fs which corresponds to the time of the sharp bend in
Fig.~\ref{fig:d64TN350_absorption}. In the DOS, the emergence of
states in the $E_g= 5.5$~eV gap of diamond can be observed. This
explains the strong increase in
absorption. Fig.~\ref{fig:d64TN350z.dos}~(c) shows the DOS at $t
= 240$~fs. At this time the intensity of the laser pulse has already
decreased significantly from its maximum value, but in the DOS,
states have completely filled the diamond gap, and the absorption of
the material is far stronger than before the pulse. 

An explanation of the overall behaviour of the absorption in the
material must be given by the time-dependent changes in the density of
states. Note, this is the main material property taken into account in
the treatment of the absorption. This time dependency of the DOS is
caused by the structural changes induced by the laser
pulse. Inspection of the trajectories shows that the sharp bend of the
graphs close to an absorbed energy of $E_0=1$~eV corresponds to the
damage threshold of the material. The damage (disorder) causes the
$E_g=5.5$~eV gap present in diamond to be filled with states until the
gap-less DOS is reached.  For pulses that are long enough to permit
the structural changes to occur before the intensity has fallen off
completely, the electronic states that fill the gap of diamond will
enormously increase the absorption of the material. Thus, the
$\tau=20$~fs graph shows only a slight increase of the absorption above
the damage threshold, because the pulse has already run out before
essential changes in the structure have taken place. On the other
hand, the $\tau=390$~fs to $\tau=500$~fs graphs show an absorption growing
sharply with pulse intensity, because all structural changes induced
by the laser can take place while the pulse intensity is still
high. Thus, the changed DOS has a strong influence on the energy that
can finally be absorbed from the pulse.

\begin{figure}
\hspace*{-0.2cm}\includegraphics[width=0.47\textwidth]{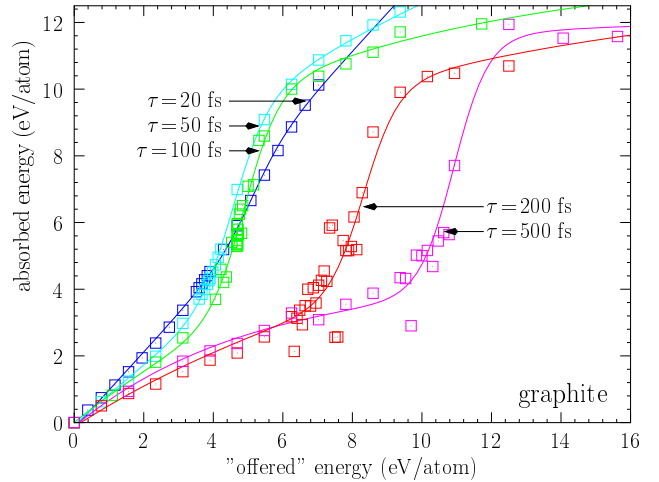}\\
\caption{
Absorbed energy per atom as a function of pulse intensity for
graphite. This is shown for a range of pulse durations from $\tau =
20$~fs to $\tau = 500$~fs. Note, the ordering of the curves with
respect to pulse duration shows a behaviour different from diamond,
which reflects a different time development of the DOS during the
pulse. Each of the results ($\boxdot$) corresponds to a calculated
trajectory. }
\label{fig:graphite_absorption}\end{figure}

In Fig.~\ref{fig:graphite_absorption}, the absorption behaviour for a
graphite sample is shown. The data are presented in the same way
as in Fig.~\ref{fig:diamond_absorption}. For graphite, the absorption
for a given pulse duration depends on the pulse intensity in a fashion
similar to that found in diamond. Again, a strong increase of
absorption takes place at the damage threshold, which in graphite is
found to be at roughly 3~eV. At this threshold, melting of graphite
and the destruction of the graphite planes takes place. At lower
intensities, absorption shows a moderate, linear increase with laser
intensity. However, an important difference can be observed in the
ordering of the curves. In the case of graphite, the slope of the
absorption curves is lower for higher pulse durations. Interpreting
this with the help of the density of states, we find that structural
relaxation taking place during the pulse decreases the density of
states close to the Fermi level and thus prevents absorption of energy
from the laser pulse. A physical picture for this effect can be
derived from the bonding character: Vibrational excitation of the
graphite planes causes atoms to be displaced perpendicular to the
plane, and thus the 3-dimensional distortion of the 2-dimensional
graphite structure causes the pure $sp^2$ bonding of graphite to be
modified with an addition of $sp^3$ bonding.

\begin{figure}
\hspace*{-0.2cm}\includegraphics[width=0.47\textwidth]{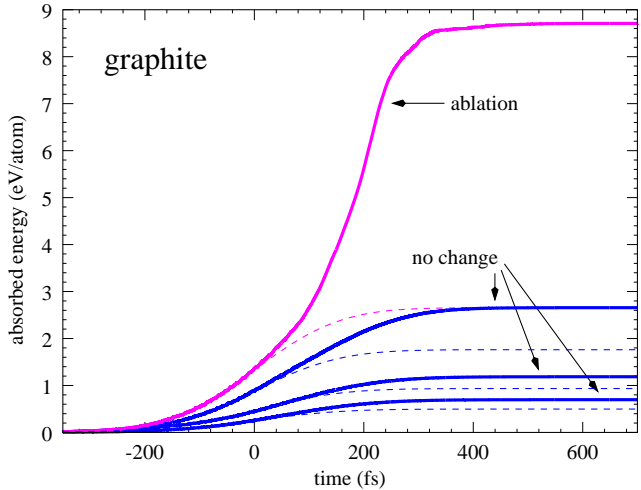}\\
\caption{
Absorbed energy per atom in graphite as a function of time for a pulse
duration of $\tau = 350$~fs. Absorption of trajectories with different
final states are shown. The dashed lines correspond to the absorption
that is expected if no structural changes occur during the action of
the pulse. }
\label{fig:c64TN350_absorption}\end{figure}

In Fig.~\ref{fig:c64TN350_absorption} we show the absorbed energy of
graphite as a function of time. The pulse duration was $\tau = 350
$~fs. Three low laser intensities produced no change in the graphite
structure, while a fourth pulse leads to ablation of the material. For
comparison, the absorbed energy that would be expected if no
structural change took place during the action of the pulse is shown
with dashed lines. As in the case of diamond (compare
Fig.~\ref{fig:d64TN350_absorption}) with increasing pulse intensity
the deviation of the actual absorbed energy from the expected value
becomes more pronounced.

\begin{figure}
\hspace*{-0.2cm}\includegraphics[width=0.47\textwidth]{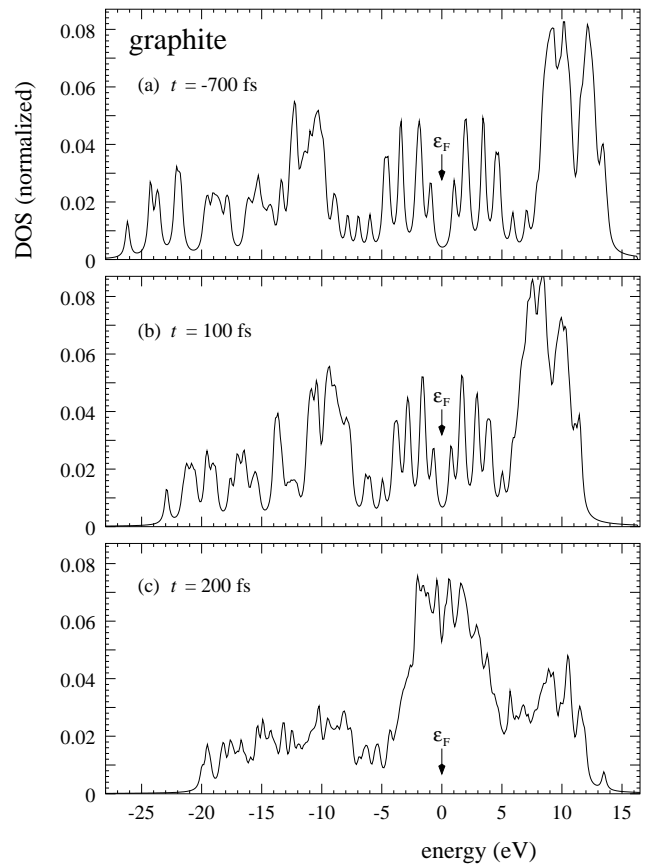}\\
\caption{
Density of states for graphite at different times for a pulse duration
of $\tau = 350$~fs. This figure corresponds to the trajectory with the
highest absorbed energy of Fig.~\ref{fig:c64TN350_absorption}. For a
discussion of the features of the DOS, see the text. }
\label{fig:c64TN350d.dos}\end{figure}

In order to analyze the cause of the increased absorption of energy we
show in Fig.~\ref{fig:c64TN350d.dos} selected densities of states
during the action of the laser pulse. These densities of states
correspond to the trajectory with the highest absorbed energy in
Fig.~\ref{fig:c64TN350_absorption}. At $t = -700$~fs, {\it i.$\:$e.} 
700~fs before the laser pulse maximum, the DOS is that of
graphite at $T = 300$~K. 
At $t = 100$~fs the overall width of
the DOS has decreased while the DOS close to the Fermi level
has increased. At $t = 200$~fs, a very high DOS around the Fermi
level has developed, indicating a metallic liquid phase of the
material. This explains the strong increase of absorption during the
action of the laser pulse. 

\begin{figure}
\hspace*{-0.2cm}\includegraphics[width=0.47\textwidth]{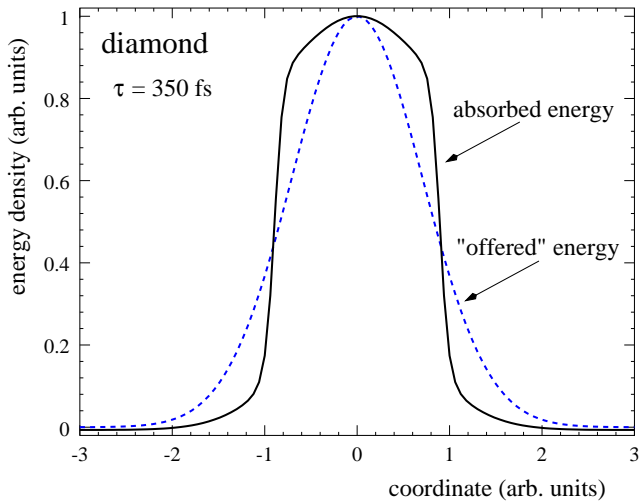}\\
\caption{
Spatial absorption profile of diamond for a pulse duration of $\tau =
350$~fs. The spatial profile of the ``offered'' laser energy is
typically given by a Gaussian (dashed line). Considering the relation
between ``offered'' energy and absorbed energy in diamond for laser
pulses of $\tau=350$~fs duration (see Fig.~\ref{fig:diamond_absorption},
we find an absorption profile that differs significantly from a
Gaussian (solid line). }
\label{fig:space_absorption}\end{figure}

In Fig.~\ref{fig:space_absorption} we show a consequence of the
behaviour of the absorbed energy for the spatial absorption profile of
diamond. Typically, a laser pulse has a Gaussian profile in space. The
different absorbed energies that result as a consequence of structural
changes in the material during the action of the pulse leads to the
spatial absorption profile shown in
Fig.~\ref{fig:space_absorption}. It is remarkable that the profile of
the absorbed energy is much steeper than the profile of the
pulse. This may be of importance in the interpretation of ablation
experiments.

\begin{figure}
\hspace*{-0.2cm}\includegraphics[width=0.47\textwidth]{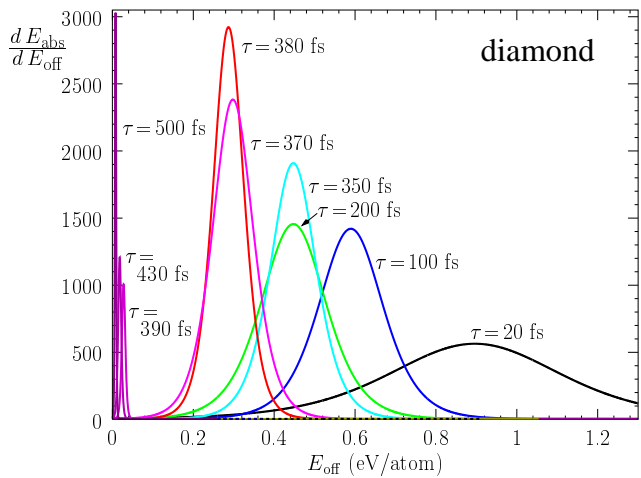}\\

\caption{
Derivative of absorbed energy with respect to ``offered'' energy in
the case of diamond for different pulse durations. The ``offered''
energy is proportional to the laser pulse intensity. The functions in
the graph are calculated as derivatives of the functions connecting
the calculated absorption values of
Fig.~\ref{fig:diamond_absorption}. Note that the functions
corresponding to all pulse durations between $\tau=20$~fs and $\tau=380$~fs
have been multiplied by a factor 30 for clarity. }
\label{fig:absorption_deriv}\end{figure}

In the introduction we have suggested characterizing the energy
absorption behavior of a system in strong nonequilibrium by a
generalized ``heat capacity'' defined as $C=dE_{\rm abs}/dI$. We
illustrate this concept in Fig.~\ref{fig:absorption_deriv} in the case
of diamond. The derivative of the absorbed energy with respect to
``offered'' energy is plotted as a function of ``offered'' energy. As
explained above we define the ``offered'' energy as the energy that
would be absorbed by the system if no structural changes occurred
during the action of the pulse. Thus, the ``offered'' energy is
proportional to the laser intensity $I$ and the quantity $dE_{\rm
abs}/dE_{\rm off}$ shown in Fig.~\ref{fig:absorption_deriv} is
proportional to $C=dE_{\rm abs}/dI$. The behavior of this quantity
shows a peak which is most pronounced for long pulse durations (see
Fig.~\ref{fig:absorption_deriv}). The peak position corresponds to the
laser intensity which induces changes in the structure and in the DOS
of the material that are most suitable to enhance the absorption of
energy. To the left of the peak, the laser is not strong enough to
induce structural changes in the material while the pulse is still
active. To the right of the peak, the laser intensity is so high that
evaporation of the material sets in during the action of the pulse
which leads to decreased absorption of energy.

\section{Summary.}\label{sec:summary}

We have calculated the absorption of energy from intense laser pulses
in diamond and graphite. An interesting nonlinear dependence of the
absorbed energy on the laser intensity is observed. Analysis of the
density of states of the materials at different times during the
action of the laser pulse shows that an enhanced absorption of energy
is due to ultrafast changes in the DOS which in turn are caused by the
structural relaxation of the excited material setting in during the
action of the exciting pulse. As these structural changes occur on a
time scale of approximately 100~fs the pulse duration plays an
important role in determining the influence of the structural
relaxation on the absorption behavior.

In summary, the results show the dependence of absorption on the light
pulse form and the occurring structural changes of these
materials. These results may also depend on the frequency and
polarisation of the laser pulse. Nevertheless, the fact that
nonthermal processes may occur during the action of the laser pulse
and thus change the absorption of the material is of general validity.

If we compare the theoretical result for the dependence of absorbed
energy on pulse intensity for graphite (see
Fig.~\ref{fig:graphite_absorption}) with the experimental result of
Ronchi {\it et~al.}~\cite{ronchi:98} 
we find a surprising similarity. Although the pulse duration
$\tau=20-30$~ms in the experiment was far higher than that considered in
the calculation, the increase of absorption at the melting point
should have a similar explanation as the increase of absorption at the
damage threshold in the calculation. It is remarkable that the effect
reported by Ronchi {\it et~al.} is also present for ultrashort laser
pulses.

\section{Acknowledgments.}

This work has been supported by the Deutsche Forschungsgemeinschaft
through SFB 450. Our simulations were done on the CRAY T3E at
Konrad-Zuse-Zentrum f{\"u}r Informationstechnik Berlin.

\end{multicols}


\begin{references}
\bibitem{sokolowski:95} 
K. Sokolowski-Tinten, J. Bialkowski and D. von der Linde,
Phys. Rev. B {\bf 51}, 14186 (1995).

\bibitem{solis:98} 
K. Sokolowski-Tinten, J. Solis, J. Bialkowski, J. Siegel, C. N. Afonso
and D. von der Linde, 
Phys. Rev. Lett. {\bf 81}, 3679 (1998).

\bibitem{jeschke:99} 
H. O. Jeschke, M. E. Garcia and K. H. Bennemann,
Phys. Rev. B {\bf 60}, R3701 (1999).

\bibitem{downer:92} 
D. H. Reitze, H. Ahn and M. C. Downer, 
Phys. Rev. {\bf B 45}, 2677 (1992). 

\bibitem{soko:98} 
K. Sokolowski-Tinten, J. Bialkowski, A. Cavalleri, D. von der Linde,
A. Oparin, J. Meyer-ter-Vehn and S. I. Anisimov, 
Phys. Rev. Lett. {\bf 81}, 224 (1998).

\bibitem{mazur:01}
J. P. Callan, A. M.-T. Kim, C. A. D. Roeser, E. Mazur, J. Solis,
J. Siegel, C. N. Afonso, and J. C. G. de Sande, 
Phys. Rev. Lett. {\bf 86}, 3650 (2001).

\bibitem{kautek:98} 
M. Lenzner, J. Kr{\"u}ger, S. Sartania, Z. Cheng, 
Ch. Spielmann, G. Mourou, W. Kautek, and F. Krausz, 
Phys. Rev. Lett. {\bf 80}, 4076 (1998).

\bibitem{rose:99} 
C. Rose-Petruck, R. Jimenez, T.  Guo, A. Cavalleri, C. W. Siders,
F. R{\'a}ksi, J. A. Squier, B. C. Walker, K. R. Wilson and
C. P. J. Barty, 
Nature {\bf 398}, 310 (1999).

\bibitem{rousse:01}
A. Rousse, C. Rischel, S. Fourmaux, I. Uschmann, S. Sebban,
G. Grillon, Ph. Balcou, E. F{\"o}rster, J. P. Geindre, P. Audebert,
J.C. Gauthier and D. Hulin, 
Nature {\bf 410}, 65 (2001).


\bibitem{parah:80} 
M. Parrinello and A. Rahman, Crystal structure and pair potentials: A
molecular-dynamics study, Phys. Rev. Lett. {\bf 45}, 1196 (1980).

\bibitem{verlet:67}
L. Verlet, Computer ``Experiments'' on Classical
Fluids. {I}. {Thermodynamical} Properties of {Lennard-Jones}
Molecules, Phys. Rev. {\bf 159}, 98 (1967).


\bibitem{haile:92}
J. M. Haile, {\it Molecular dynamics simulation: elementary methods},
John Wiley \& Sons, New York 1975.


\bibitem{ronchi:98} M. Musella, C. Ronchi, M. Brykin and 
M. Sheindlin, J. App. Phys. {\bf 84}, 2530 (1998).

\end{references}
\end{document}